%% file: Template.tex
\documentclass{article}
\usepackage{spconf,amsmath,amssymb,graphicx,hyperref}

\makeatletter
\def\section{\@startsection{section}{1}{\z@}{-10pt}{6pt}{}} %default ~14, ~9
\def\subsection{\@startsection{subsection}{2}{\z@}{-10pt}{4pt}{}} %default ~13, ~6
\makeatother

\usepackage{enumitem}
\usepackage{tikz}
\usetikzlibrary{arrows.meta,decorations.pathreplacing}

\def\x{{\mathbf x}}

\title{Speech Block Influence: Component-Specific Layer Scoring \\ for Pruning Speech LLMs}
\name{Siyu Yao  \quad Du Q.~Huynh \quad Lian Xu \quad Mark Reynolds}
\address{%School of Physics, Mathematics and Computing\\Department of Computer Science and Software Engineering\\
The University of Western Australia \\
siyu.yao@research.uwa.edu.au; \{du.huynh, lian.xu, mark.reynolds\}@uwa.edu.au}
\begin{document}
\ninept
\maketitle

\newcommand{\SBIenc}{\textnormal{SBI-Enc}}
\newcommand{\SBIdec}{\textnormal{SBI-Dec}}
\newcommand{\BIT}{\mathrm{BI}^{\mathcal{T}}}
\newcommand{\BIA}{\mathrm{BI}^{\mathcal{A}}}
\newcommand{\BIAT}{\mathrm{BI}^{\mathcal{A\cup T}}}
\newcommand{\BItext}{\mathrm{BI}^{\mathrm{text}}}

\begin{abstract}
% The abstract should appear at the top of the left-hand column of text, about
% 0.5 inch (12 mm) below the title area and no more than 3.125 inches (80 mm) in
% length.  Leave a 0.5 inch (12 mm) space between the end of the abstract and the
% beginning of the main text.  The abstract should contain about 100 to 150
% words, and should be identical to the abstract text submitted electronically
% along with the paper cover sheet.  All manuscripts must be in English, printed
% in black ink.

Speech LLMs are costly to deploy in resource-constrained settings. Layer pruning can cut this cost, but existing scoring metrics transfer poorly to speech LLMs: they assume a decoder-only architecture with homogeneous token sequences, whereas speech LLMs add encoder and adapter components and process multimodal sequences. We propose \textbf{Speech Block Influence} (\textbf{SBI}), the first layer-importance scoring framework designed for speech LLM pruning that consists of two component-specific scores: \textbf{SBI-Enc} measures the effect of encoder-layer removal at the adapter's output to better reflect downstream impact; \textbf{SBI-Dec} measures layer-wise input–output similarity over text-token positions only to avoid audio-token dominance. Across three speech LLMs, SBI improves pruning robustness, with stronger encoder performance at higher pruning rates and more reliable decoder layer selection by scoring text tokens rather than the audio-dominated full sequence. We further find that text-only calibration yields decoder rankings highly correlated with those from speech-text calibration, suggesting a cheaper alternative to measure decoder layer importance. 

\end{abstract}
\section{Introduction}
\label{sec:intro}

% These guidelines include complete descriptions of the fonts, spacing, and
% related information for producing your proceedings manuscripts. Please follow
% them and if you have any questions, direct them to Conference Management
% Services, Inc.: Phone +1-979-846-6800 or email
% to \\\texttt{papers@2027.ieeeicassp.org}.

Speech large language models (speech LLMs) bring general language understanding to spoken interfaces such as voice assistants. Although speech LLMs achieve strong speech understanding~\cite{chu24qwen2audio, goel25audioflamingo3, liu25voxtral}, they are too large to deploy on resource-constrained devices such as phones and in-car systems. These models typically consist of three components~\cite{peng26speechLLM_understanding}: a speech encoder that converts input audio into speech representations; an adapter that projects them into the LLM embedding space; and a large decoder-only transformer LLM that processes these representations and generates text. Each transformer layer in the encoder and the decoder increases the memory footprint and the latency of every forward pass; compressing both components is therefore necessary for efficient deployment.

One way to reduce the computational demand is \textit{layer pruning}. Using a small set of unlabelled inputs referred to as the \textit{calibration set}, this pruning strategy identifies and removes redundant transformer layers of the model.
Studies of model pruning show that a large fraction of text LLM parameters can be removed with little loss in downstream task accuracy~\cite{frantar23sparsegpt, gromov25unreasonable, men25shortgpt}; speech LLMs built on these backbones may inherit this redundant capacity. However, pruning remains underexplored in speech LLMs, which differ from text LLMs in two ways: they include additional speech encoder and adapter components, and their input sequence contains audio embeddings in addition to text. ShortGPT~\cite{men25shortgpt}, a strong and representative baseline for training-free LLM layer pruning, scores each layer by its Block Influence (BI) and removes the least important layers. This criterion was able to retain 90\% of a text LLM's performance while removing 25\% of the parameters without healing. However, we show that BI does not transfer effectively to speech LLM pruning.

We present Speech Block Influence (SBI), the first layer-importance scoring framework designed for speech LLM pruning. Our contributions are summarised below:
\begin{enumerate}[itemsep=0.5pt,topsep=2pt]
    \item We present the first systematic analysis of layer redundancy across both the encoder and decoder of speech LLMs, showing that the BI score can be suboptimal.
    \item We propose SBI with component-specific scoring for speech LLMs: SBI-Enc measures the effect of encoder-layer removal at the adapter's output, while SBI-Dec scores decoder layers over text tokens only, improving pruning robustness over BI.
    \item We show that rankings from text-only calibration closely correlate with those from speech-text calibration, suggesting a cheaper alternative for estimating decoder layer importance.
\end{enumerate}

\section{Related Work}
\label{sec:lit-rev}

Model pruning ranges from unstructured, at the level of individual weights~\cite{frankle18lottery,frantar23sparsegpt, sun24wanda}, to structured, which removes entire units such as attention heads and MLP channels~\cite{ma23llmpruner, wang26gisp}. Layer pruning is a special case of structured pruning where entire transformer blocks are removed. Recent studies show that LLMs carry many redundant layers~\cite{men25shortgpt,gromov25unreasonable,kim2024shortenedllama, yang24laco}. ShortGPT~\cite{men25shortgpt} introduces the BI score, which measures how much a layer changes its hidden states. A layer that barely changes the hidden states is assumed to contribute little. The BI score for layer $\ell$ is given by:
{
\setlength{\abovedisplayskip}{1pt}
\setlength{\belowdisplayskip}{1pt}
\begin{equation}
\label{eq:bi}
\mathrm{BI}_\ell \;\triangleq\; 1 \;-\; \mathbb{E}_{n,t}\!\left[
\frac{\mathbf{h}_{\ell,t}^{(n)\top}\,\mathbf{h}_{\ell+1,t}^{(n)}}
     {\lVert \mathbf{h}_{\ell,t}^{(n)} \rVert_2 \,
      \lVert \mathbf{h}_{\ell+1,t}^{(n)} \rVert_2}\right],
\end{equation}
}%
where $\mathbf{h}_{\ell,t}^{(n)}$ denotes the input hidden state of layer $\ell$ at the $t^{\text{th}}$ token of the $n^{\text{th}}$ calibration sample. Concurrently, Gromov et al.~\cite{gromov25unreasonable} instead measure the angular distance between the input and output of contiguous blocks of layers to identify and remove the most redundant block. Their results show that deeper layers are more redundant than shallow layers in text LLMs and can be pruned with minimal loss.

 \begin{figure*}[t]
\centering
\begin{minipage}[b]{0.5\textwidth}
  \centering
  \input{sbi_enc}
  \centerline{(a) SBI-Enc: encoder scoring}
\end{minipage}
\hfill
\begin{minipage}[b]{0.47\textwidth}
  \centering
  \input{sbi_dec}
  \centerline{(b) SBI-Dec: decoder scoring}
\end{minipage}
\vspace{-6pt}
\caption{The proposed SBI framework for speech LLM layer scoring. (a) SBI-Enc: the calibration input $\mathbf{x}^{(n)}$ is passed through the full encoder and the encoder with layer $\ell$ removed; the resulting adapter outputs are compared to obtain $\SBIenc_\ell$. (b) SBI-Dec: the input and output text tokens of decoder layer $\ell$ are compared to obtain $\SBIdec_\ell$. In contrast, $\mathrm{BI}^{\mathcal{A}}$ is computed over audio tokens and $\mathrm{BI}^{\mathcal{A}\cup\mathcal{T}}$ over all tokens.
}
\label{fig:sbi}
\end{figure*}
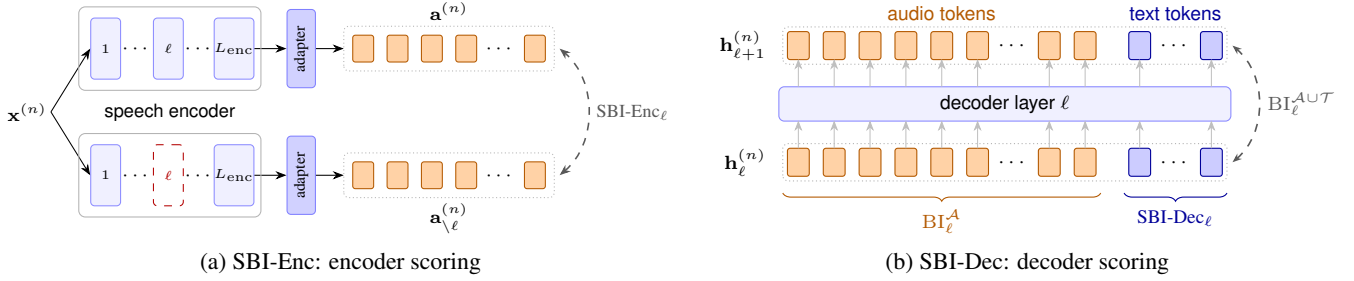 

Most existing speech LLM compression studies focus only on retaining the automatic speech translation (AST) performance of the model~\cite{moslem25efficient_compression,liu26dietkit,palomino26codec_compression}. Two recent studies prune speech LLMs built with the SLAM recipe~\cite{ma2024slam}: a frozen speech encoder and LLM decoder with only the adapter trained. Moumen et al.~\cite{moumen26speechllm_decoder_redundancy} prune contiguous blocks of decoder layers using the angular distance metric~\cite{gromov25unreasonable}. Kolluri et al.~\cite{kolluri26encoder_pruning}, in contrast, prune the encoder, removing the deepest Whisper layers and recovering performance with LoRA adapters on the LLM. All prior work evaluates the pruned models on automatic speech recognition (ASR) and AST only. Our study differs from previous speech LLM pruning studies in three ways:
we propose a scoring framework specifically for speech LLM layers; we study natively trained speech LLMs, whose decoders have been fine-tuned on speech data; and we evaluate broader speech understanding with multiple-choice questions alongside ASR.

\section{The SBI Framework}
\label{sec:method}
% The paper title (on the first page) should begin 1.38 inches (35 mm) from the
% top edge of the page, centered, completely capitalized, and in Times 14-point,
% boldface type.  The authors' name(s) and affiliation(s) appear below the title
% in capital and lower case letters.  Papers with multiple authors and
% affiliations may require two or more lines for this information. Please note
% that papers should not be submitted blind; include the authors' names on the
% PDF.

SBI scores encoder and decoder layers separately: for the encoder, it measures the change in the adapter's output after removing each layer (Section~\ref{sec:enc}); for the decoder, it measures each layer's input-output similarity over text tokens only  (Section~\ref{sec:dec}). Sections~\ref{sec:calib} and~\ref{sec:prune} describe the calibration set and pruning procedure.

\subsection{Encoder layer scoring at the adapter's output}
\label{sec:enc}
A speech LLM's decoder does not consume the encoder output directly: the adapter projects and downsamples the encoder's output. A large change to the hidden states by an encoder layer may have a small effect on the adapter's output, so the layer would contribute little to downstream computation; however, scores that compare a layer's input and output, such as BI~\cite{men25shortgpt} and angular distance~\cite{gromov25unreasonable}, would still score this layer as important. As the internal hidden states can be a misleading proxy for redundancy, we propose to score encoder layers at the adapter's output instead. Let $\mathbf{a}^{(n)}$ be the output embedding from the adapter for the unpruned model with the $n^{\text{th}}$ calibration sample as input, and $\mathbf{a}^{(n)}_{\setminus \ell}$ be the adapter's output of the same model with the $\ell^{\text{th}}$ encoder layer removed. We define the encoder score ($\SBIenc$) of the $\ell^{\text{th}}$ layer, illustrated in Fig.~\ref{fig:sbi}(a), as:
{
\setlength{\abovedisplayskip}{0pt}
\setlength{\belowdisplayskip}{0pt}
\begin{equation}
\label{eq:ao}
% \text{SBI-Enc}(l) \equiv
\SBIenc_{\ell} \,\triangleq\, 1 \;-\; \mathbb{E}_{n,t}\!\left[
\frac{\mathbf{a}^{(n)\top}_{t}\,\mathbf{a}^{(n)}_{\setminus \ell,\,t}}
     {\lVert \mathbf{a}^{(n)}_{t} \rVert_2\,
      \lVert \mathbf{a}^{(n)}_{\setminus \ell,\,t} \rVert_2}\right],
      %,
%\; \text{for } \ell = 1,\cdots,L_{\mathrm{Enc}},
\end{equation}
}
where $\mathbf{a}^{(n)}_{t}$ denotes the adapter's output at token position $t$.
A low $\SBIenc_\ell$ means the adapter's output is nearly unchanged when the $\ell^{\text{th}}$ layer is removed, i.e., the layer is redundant. 

\subsection{Decoder layer scoring with text tokens}
\label{sec:dec}
In a speech LLM, the decoder's input sequence is not homogeneous: it comprises the audio embeddings produced by the encoder and adapter, concatenated with the text embeddings of the prompt. In our calibration set, audio tokens outnumber text tokens by $6\times$ to $22\times$ depending on the model's encoder token rate and padding. Pruning according to the audio-token-dominated score collapses the speech LLM after removing as few as one layer in some models (Fig.~\ref{fig:pruning}(b)). We therefore partition the tokens into an audio set $\mathcal{A}$ and a text set $\mathcal{T}$, and define the score of the $\ell^{\text{th}}$ decoder layer for text tokens as:
{
\setlength{\abovedisplayskip}{0pt}
\setlength{\belowdisplayskip}{0pt}
\begin{equation}
\label{eq:bi-pos}
% \text{SBI-Dec}(l) \!\equiv
\SBIdec_{\ell} \triangleq 1 \,-\, \mathbb{E}_{n,\,t\in\mathcal{T}^{(n)}}\!\left[
\frac{\mathbf{h}_{\ell,t}^{(n)\top}\,\mathbf{h}_{\ell+1,t}^{(n)}}
     {\lVert \mathbf{h}_{\ell,t}^{(n)} \rVert_2\,
      \lVert \mathbf{h}_{\ell+1,t}^{(n)} \rVert_2}\right],
\end{equation}
}
where $\mathcal{T}^{(n)}$ is the set of text-token positions in sample $n$. $\SBIdec$ keeps the original BI formulation but applies it only to text tokens (Fig.~\ref{fig:sbi}(b)). Replacing $\mathcal{T}$ by $\mathcal{A}$ or $\mathcal{A}\cup\mathcal{T}$ in Eq.~\eqref{eq:bi-pos} gives $\BIA$ or $\BIAT$, where $\BIAT$ is the original BI given in Eq.~\eqref{eq:bi}.

\begin{figure*}[tp!]
  \centering
  \includegraphics[width=0.94\textwidth]{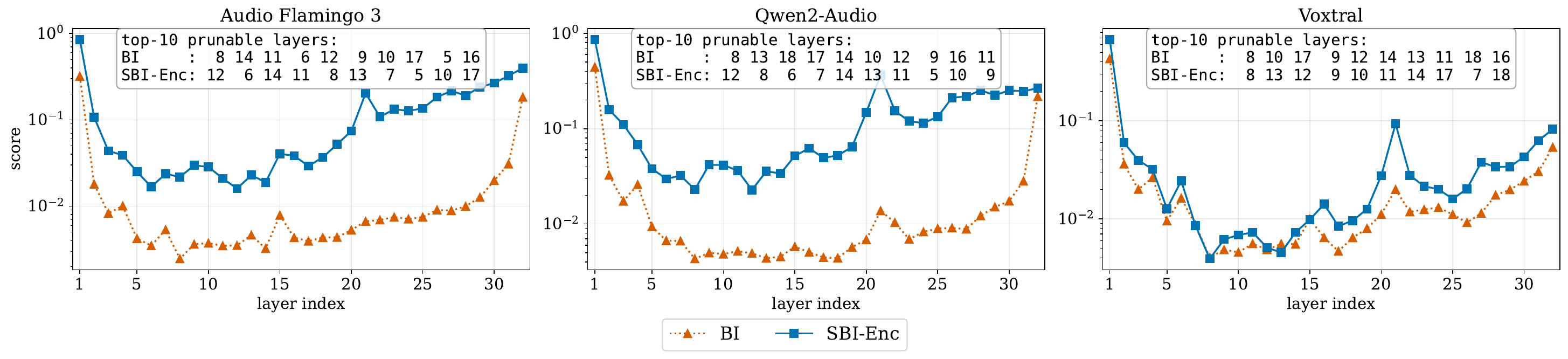}

  \includegraphics[width=0.94\textwidth]{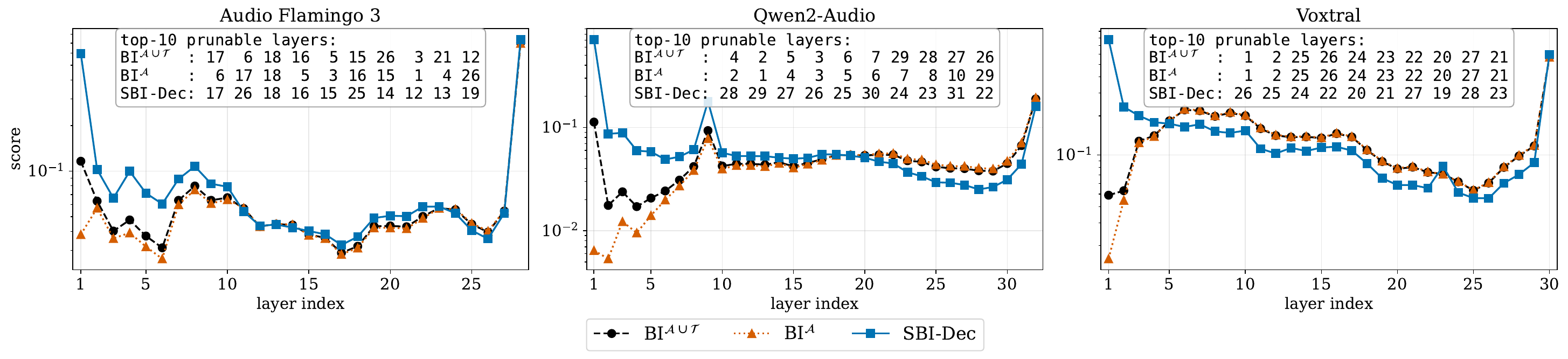}
  \\[-8pt]
  \caption{Layer importance scores in log scale for the three speech LLMs, where lower values indicate higher layer redundancy. Top row: encoder pruning with $\SBIenc$ versus original BI. Bottom row: decoder pruning with $\SBIdec$ versus $\mathrm{BI}$ computed over all tokens $\BIAT$ and audio tokens $\BIA$. Insets list the top-10 prunable layers selected by each score, in pruning order.
  \\[-10pt]}
  \label{fig:importance}
\end{figure*}

 \subsection{Calibration set}
%\subsection{Calibration set for measuring $\BIT$, $\BIA$, $\BIAT$, and $\BItext$}
\label{sec:calib}
To compute layer importance scores, we construct a speech-text calibration set. Ji et al.~\cite{ji25calibration} propose synthesising calibration data with the model itself, after finding that data resembling the pre-training distribution works best. The speech side cannot be synthesised this way, as the speech LLMs that we evaluate cannot generate speech. We therefore sample 30 audio clips each from Common Voice~\cite{ardila20commonvoice}, IEMOCAP~\cite{busso08iemocap}, VoxCeleb~\cite{NAGRANI20voxceleb}, and Gaokao~\cite{hu24wavllm}. These four corpora were chosen for their variation in accent, emotion, number of speakers, and clip length, %so that scores are measured across
giving diverse acoustic conditions for the measured scores. The text prompt for each clip is generated by the model, using the instruction \textit{``Listen to the audio and generate one question about the speech''}. We select 100 examples having the lowest generation perplexity as that model's calibration set.
Calibration clips come from separate splits and are not used for evaluation.

Although $\SBIdec$ excludes audio tokens, the audio context is still available for the model to attend to. Whether the resulting ranking depends on the audio at all is therefore not obvious.  We test this by comparing $\SBIdec$, computed on the speech-text calibration set mentioned above, against $\mathrm{BI}^{\text{text}}$, the original BI score of Eq.~\eqref{eq:bi} computed on a text-only calibration set. We construct this text-only set by transcribing the speech in every speech-text calibration example to ensure the two sets only differ in modality, not in content. 

\subsection{Pruning procedure}
\label{sec:prune}
Layer importance scores are computed once on the unpruned model, and the $k$ lowest-scoring layers are removed. The encoder and decoder are pruned separately, so each comparison isolates the selection criterion for one component. No healing is applied after pruning, as this is a controlled comparison of layer selection criteria for post-training pruning. Computing $\SBIenc$ requires $L_{\mathrm{Enc}}+1$ passes over the calibration set, whereas $\SBIdec$ requires a single pass. 

\section{Experimental Setup}
\label{sec:setup}
\textbf{Models.}
We evaluate pruning on three speech LLMs with different decoder backbones: Qwen2-Audio-7B-Instruct~\cite{chu24qwen2audio} (Qwen-7B decoder, $L_{\mathrm{Dec}}\!=\!32$), Audio Flamingo 3~\cite{goel25audioflamingo3} (Qwen2.5-7B decoder, $L_{\mathrm{Dec}}\!=\!28$), and Voxtral Mini 3B~\cite{liu25voxtral} (Ministral-3B decoder, $L_{\mathrm{Dec}}\!=\!30$). All of them use a 32-layer encoder ($L_{\mathrm{Enc}}\!=\!32$) initialised from the Whisper-large-v3~\cite{radford23whisper} encoder, but have different audio token rates: %. Audio token rate differs across models; 
Voxtral has a token rate of 12.5\,Hz with padding to a multiple of 30\,s, while the other two models have a 25\,Hz token rate. 
\\[2pt]
\textbf{Tasks and evaluation measures.} The two tasks being investigated are ASR and multiple-choice question (MCQ) answering. For ASR, word error rate (WER) is the evaluation measure; for MCQ, each question has $4$ possible choices, and accuracy is used for evaluation. WER above 100\% reflects degenerate repetition, and MCQ accuracy below random guessing (25\%) indicates that the model no longer emits a valid answer. We treat these cases as model collapse, beyond which comparisons between criteria are not meaningful.
\\[2pt]
\textbf{Benchmark datasets.}
The pruned models are evaluated on three benchmark datasets: Common Voice~\cite{ardila20commonvoice}, 
MMSU~\cite{wang26mmsu}, and the OpenBookQA (OBQA) subset of VoiceBench~\cite{chen26voicebench}.
Common Voice is used for the ASR task. We randomly sample $100$ examples for evaluation.
The last two datasets are for the MCQ task.
Each MMSU item is a speech clip paired with a text question targeting one of four categories: semantics, phonology, speaker traits and speaking style. In OBQA, the question is presented as speech, and only linguistic understanding is tested. In MMSU, we randomly sample 125 examples from each of the four categories, giving 500 questions; for OBQA we use the full dataset of 455 questions.
\\[2pt]
\textbf{Baselines.}
We prune up to 10 layers from the encoder and decoder of each speech LLM independently. We compare $\SBIenc$ with BI for encoder pruning and $\SBIdec$ with $\mathrm{BI}^{\mathcal{A}\cup\mathcal{T}}$ for decoder pruning. 

\begin{figure*}[tp!]
  \centering
  \includegraphics[width=0.96\textwidth]{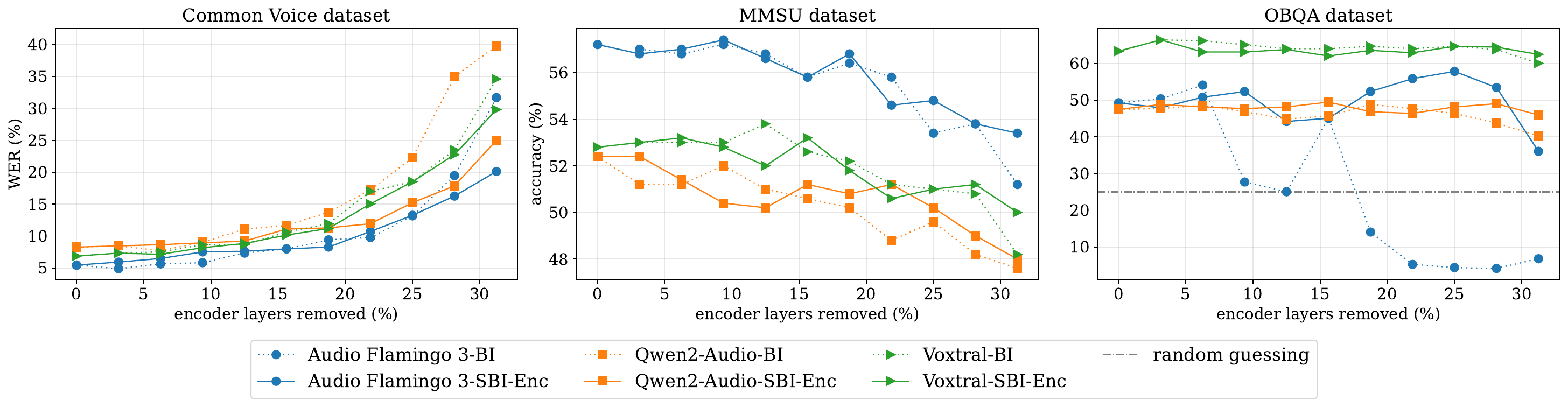}
  \centerline{(a) Encoder pruning: $\SBIenc$ versus $\mathrm{BI}$}

  \includegraphics[width=0.96\textwidth]{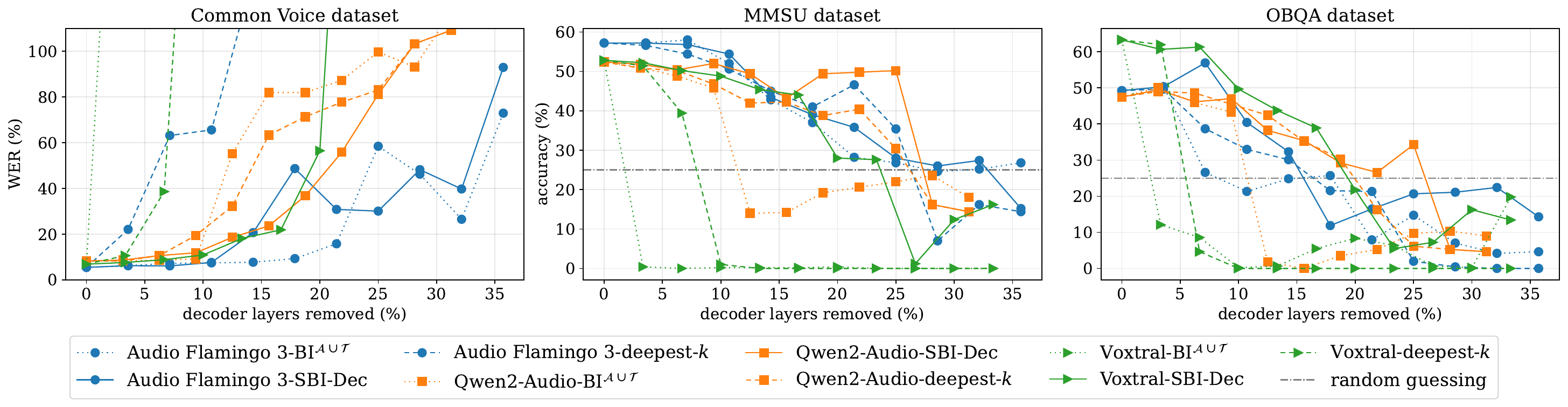}
  \centerline{(b) Decoder pruning: $\SBIdec$ versus $\mathrm{BI}^{\mathcal{A}\cup\mathcal{T}}$ and deepest-$k$.}
  \caption{Model performance after pruning $k\in[0,10]$ layers (up to $31\%$ for the encoder and $31$-$36\%$ for the decoder). Solid lines: proposed SBI; dotted lines: BI; dashed lines (decoder only): deepest-$k$. The horizontal lines mark the 25\% random guessing baseline for MCQ tasks.}
  \label{fig:pruning}
\end{figure*}

\section{Results and Discussion}
\label{sec:results}
Fig.~\ref{fig:importance} shows the layer importance scores and the top 10 prunable layers for all three models. 
$\SBIenc$ and $\mathrm{BI}$ select largely the same layers but in different orders: their top-10 sets share seven to nine of ten layers across the three models. Where they differ, $\SBIenc$ selects shallower layers and has a lower mean top-10 index in every model. 
In the decoders, the audio token dominance causes $\mathrm{BI}^{\mathcal{A}\cup\mathcal{T}}$ to closely follow $\mathrm{BI}^{\mathcal{A}}$, especially for Voxtral, where the two select an identical set of layers. $\mathrm{BI}^{\mathcal{A}}$ always selects some shallow layers %with mean indices of 10.1, 6.5 and 18.1, 
while $\SBIdec$ selects deeper layers, never below index 10, consistent with Gromov et al.'s~\cite{gromov25unreasonable} finding that deeper layers are the most redundant. To test whether the performance of $\SBIdec$ comes from layer depth, we additionally test deepest-$k$ pruning, which removes the $k$ deepest decoder layers (excluding the last layer, as it has a high score in all three models). Code is available at \url{https://github.com/CU-0/sbi-speechllm-pruning}.
%\\[2pt]
%
\subsection{Encoder pruning}
% \label{sec:encoder}
%\textbf{Encoder Pruning} \\
Fig.~\ref{fig:pruning}(a) reports the performance as encoder layers are pruned according to $\SBIenc$ versus $\mathrm{BI}$. Pruning layers of the encoder degrades the performance far less than pruning the same number of decoder layers. Encoder layers are also much smaller than decoder layers, e.g., Qwen2-Audio's decoder has %is
roughly 11$\times$ the number of parameters %size 
of its encoder. 
When fewer than 20\% of encoder layers are removed, the curves generally overlap and cross, with the exception of Audio Flamingo 3 on OBQA, suggesting that neither criterion has a consistent advantage. 
The difference emerges at higher pruning rates, where $\SBIenc$ matches or outperforms $\mathrm{BI}$ in every model when 31\% of layers are removed. 
\\
\indent For the Common Voice dataset, $\SBIenc$ gives the lowest WER for the ASR task %preserves ASR better 
in all three models. The gap is largest for Qwen2-Audio (orange curves), where removing 31\% of encoder layers gives 40\% WER under $\mathrm{BI}$ against 25\% under $\SBIenc$. The most prominent difference is for Audio Flamingo 3 on the OBQA dataset: pruning by $\mathrm{BI}$ falls below random guessing at 19\% of removed layers (blue dotted curve), while $\SBIenc$ maintains good accuracy until pruning 28\% of encoder layers. For MMSU, the two criteria are weakly separated and the curves overlap most of the time. Overall, $\SBIenc$ better preserves performance for Qwen2-Audio and Audio Flamingo 3, but the difference is minor for Voxtral, as the two criteria share nine layers in their top-10 pruning set (Fig.~\ref{fig:importance}). 
%\\[2pt]
\subsection{Decoder pruning}
 \label{sec:decoder}
%\textbf{Decoder Pruning} \\
Fig.~\ref{fig:pruning}(b) compares decoder pruning with $\SBIdec$, $\BIAT$, and deepest-$k$. 
Results suggest that $\SBIdec$ is the better criterion for all three models on almost every task. This is the opposite of the trend reported for vision-language models, where scoring on text tokens alone performs worse than scoring on visual or all tokens~\cite{ma25shortlvlm}. One exception is the ASR performance of Audio Flamingo 3, where $\BIAT$ gives lower WER than $\SBIdec$ for most pruning rates. Pruning by $\SBIdec$ consistently delays collapse: for the MCQ task for both MMSU and OBQA datasets, $\BIAT$ falls below random guessing at a smaller pruning rate than $\SBIdec$ in every model.  

Although $\SBIdec$ favours deep layers, it is not equivalent to removing the deepest layers. Deepest-$k$ raises WER faster than $\SBIdec$ in all three models and collapses Voxtral after two to three layers (7–10\%). On the MCQ tasks, $\SBIdec$ is the better criterion, except for Audio Flamingo 3 on MMSU. For Qwen2-Audio, even though $\SBIdec$ and deepest-$k$ share the same top-10 set, the ordering from $\SBIdec$ still preserves performance better up to 25\% of layers pruned. Even under $\SBIdec$, speech LLMs generally degrade substantially before 20\% of decoder layers are removed. In contrast, text LLM pruning studies report minimal degradation of performance at 25\% to 30\% of layer pruning~\cite{gromov25unreasonable, men25shortgpt}. This suggests that multimodal decoders may be more sensitive to layer pruning, which is similarly observed in pruning vision-language models~\cite{ma25shortlvlm}.

\subsection{Text-only calibration}
\label{sec:calibration}
%\textbf{Text-only calibration} \\
Table~\ref{tab:overlap} compares $\SBIdec$ with $\mathrm{BI}^{\text{text}}$, the BI score given in Eq.~\eqref{eq:bi} computed on the text-only calibration set (Section~\ref{sec:calib}). The two scores rank decoder layers almost identically: Spearman's $\rho \ge 0.93$ for all models. The agreement also holds for the pruned layer sets. For $k \in \{3, 6, 9, 12\}$, the $k$ lowest-scoring layer sets are identical in eight of the twelve settings and never differ by more than one layer. Therefore, scoring over text tokens selects nearly the same layers whether or not the calibration data contains speech. As audio occupies the majority of the multimodal sequence, the text-only pass uses far fewer tokens and skips the encoder entirely, making it a cheaper alternative for measuring decoder layer importance.

\begin{table}[t!]
\centering
\caption{Top-$k$ overlap ($i/j$ layers shared) and Spearman's $\rho$ between $\SBIdec$ and $\mathrm{BI}^{\text{text}}$ decoder rankings.}
\label{tab:overlap}
{\footnotesize
\begin{tabular}{lcccc|c}
\hline
Model & $k=3$ & $k=6$ & $k=9$ & $k=12$ & $\rho$ \\
\hline
Audio Flamingo 3 & 3/3 & 6/6 & 8/9 & 11/12 & 0.99 \\
Qwen2-Audio      & 2/3 & 6/6 & 9/9 & 12/12 & 0.93 \\
Voxtral          & 3/3 & 6/6 & 8/9 & 12/12 & 0.99 \\
\hline
\end{tabular}
}
\end{table}

\section{Conclusion and future work}
\label{sec:conc}
We have presented SBI, a layer-importance scoring framework for pruning both the encoder and the decoder of speech LLMs. SBI-Enc scores encoder layers by their effect of layer removal on the adapter's output, while SBI-Dec measures decoder-layer importance over text tokens only. Across three speech LLMs, SBI improves pruning robustness over the original BI score in most settings. We also found that scores computed on text-only calibration data correlate closely with those computed on speech-text data, making text-only calibration a cheaper alternative for estimating decoder layer importance. 
Two directions remain for future work. First, all results are reported without healing; combining lightweight fine-tuning such as LoRA with SBI may allow higher pruning rates. Second, we prune the encoder and decoder independently; jointly pruning these components may reach a better compression-performance trade-off.

\vfill\pagebreak

% \section{REFERENCES}
% \label{sec:refs}

% List and number all bibliographical references at the end of the
% paper. The references can be numbered in alphabetic order or in
% order of appearance in the document. When referring to them in
% the text, type the corresponding reference number in square
% brackets as shown at the end of this sentence \cite{C2}. An
% additional final page (the fifth page, in most cases) is
% allowed, but must contain only references to the prior
% literature.

% Please follow the IEEE Citation Guidelines, \url{https://ieee-dataport.org/sites/default/files/analysis/27/IEEE\%20Citation\%20Guidelines.pdf} for formatting of references.

% References should be produced using the bibtex program from suitable
% BiBTeX files (here: strings, refs, manuals). The IEEEbib.bst bibliography
% style file from IEEE produces unsorted bibliography list.
% -------------------------------------------------------------------------

\clearpage

\section{Compliance with Ethical Standards}
\label{sec:ethics}
This study was conducted using publicly released speech corpora. No new data were collected from human subjects, and no ethical approval was required for this study.

\section{Acknowledgements}
\label{sec:ack}
This research was supported by an Australian Government Research Training Program (RTP) Scholarship.

\bibliographystyle{IEEEbib}
\bibliography{main}

\end{document}

%% file: sbi_enc.tex
\resizebox{\linewidth}{!}{%
\begin{tikzpicture}[
  font=\sffamily,
  lay/.style={draw, rounded corners=1.5pt, minimum width=0.40cm,
              minimum height=0.82cm, inner sep=0pt, fill=blue!6, draw=blue!45,
              font=\sffamily\tiny},
  layx/.style={lay, dashed, fill=white, draw=red!70!black, text=red!70!black},
  adp/.style={draw, rounded corners=1.5pt, minimum width=0.38cm,
              minimum height=1.05cm, inner sep=0pt, fill=blue!18, draw=blue!50,
              font=\fontsize{6}{7}\selectfont},
  inp/.style={minimum width=0.62cm, minimum height=0.50cm, inner sep=1pt,
              font=\sffamily\scriptsize},
  aud/.style={rounded corners=1pt, minimum width=0.28cm,
              minimum height=0.36cm, inner sep=0pt,
              fill=orange!35, draw=orange!70!black},
  lbl/.style={font=\sffamily\scriptsize},
  dts/.style={font=\scriptsize, inner sep=0pt},
  frm/.style={draw=gray!60, densely dotted, rounded corners=2pt},
  cmp/.style={<->, semithick, gray!70!black, dashed},
  >={Stealth[length=1.5mm]}
]
\def\dx{0.46}    % adapter-output token spacing
\def\xt{3.45}    % first adapter-output token
\node[inp] (xin) at (-1.05,0.70) {$\mathbf{x}^{(n)}$};
\foreach \row/\yy/\sty in {u/1.55/lay, p/-0.15/layx}{
  \node[lay]  (\row L1) at (0.00,\yy) {$1$};
  \node[dts]  at (0.42,\yy) {$\cdots$};
  \node[\sty] (\row L2) at (0.84,\yy) {$\ell$};
  \node[dts]  at (1.26,\yy) {$\cdots$};
  \node[lay]  (\row L3) at (1.72,\yy) {$L_{\mathrm{enc}}$};
  \node[adp]  (\row ad) at (2.62,\yy) {\rotatebox{90}{adapter}};
  \draw[->] (\row L3.east) -- (\row ad.west);
  \draw[->] (xin.east) -- (\row L1.west);
}
%%% adapter outputs, framed
\foreach \yy/\r in {1.55/u, -0.15/p}{
  \foreach \i in {0,1,2,3,5}{ \node[aud] at (\xt+\i*\dx,\yy) {}; }
  \node[dts] at (\xt+4*\dx,\yy) {$\cdots$};
  \draw[frm] (\xt-0.26,\yy-0.30) rectangle (\xt+5*\dx+0.26,\yy+0.30);
  \coordinate (\r-fr) at (\xt+5*\dx+0.26,\yy);
  \draw[->] (\r ad.east) -- (\xt-0.26,\yy);
}
%%% encoder boundary + labels
\foreach \yy in {1.55,-0.15}{
  \draw[rounded corners=2.5pt, gray!60]
        (-0.36,\yy-0.56) rectangle (2.08,\yy+0.56);}
\node[lbl] at (0.86,0.70) {speech encoder};
\node[lbl] at (\xt+2.5*\dx,1.55+0.50) {$\mathbf{a}^{(n)}$};
\node[lbl] at (\xt+2.5*\dx,-0.15-0.58) {$\mathbf{a}^{(n)}_{\setminus\ell}$};
%%% AI compares the two adapter outputs position by position
\draw[cmp] ([xshift=2pt]p-fr) to[bend right=50]
  node[midway, right, lbl] {$\SBIenc_\ell$} ([xshift=2pt]u-fr);
\end{tikzpicture}}

%% file: sbi_dec.tex
\resizebox{\linewidth}{!}{%
\begin{tikzpicture}[
  font=\sffamily,
  tok/.style={rounded corners=1pt, minimum width=0.28cm,
              minimum height=0.36cm, inner sep=0pt},
  aud/.style={tok, fill=orange!35, draw=orange!70!black},
  txt/.style={tok, fill=blue!20, draw=blue!60!black},
  lay/.style={draw, rounded corners=2pt, fill=blue!6, draw=blue!45,
              minimum height=0.44cm, font=\sffamily\scriptsize},
  lbl/.style={font=\sffamily\scriptsize},
  dts/.style={font=\scriptsize, inner sep=0pt},
  brc/.style={decorate, decoration={brace, amplitude=3pt, raise=1pt}},
  frm/.style={draw=gray!60, densely dotted, rounded corners=2pt},
  flw/.style={->, very thin, gray!50},
  cmp/.style={<->, semithick, gray!70!black, dashed},
  >={Stealth[length=1.2mm]}
]
\def\dx{0.46}      % token spacing
\def\gap{0.24}     % extra space between audio and text
\def\yin{0}        % layer input row
\def\yout{1.5}     % layer output row
\def\ylay{0.75}    % layer centre
% x positions: audio slots 0..8 (slot 6 = dots), text slots 9..11 (slot 10 = dots)
\pgfmathsetmacro{\xAl}{0}               % first audio
\pgfmathsetmacro{\xAr}{8*\dx}           % last audio
\pgfmathsetmacro{\xTl}{9*\dx+\gap}      % first text
\pgfmathsetmacro{\xTr}{11*\dx+\gap}     % last text
\pgfmathsetmacro{\xMid}{(\xAl+\xTr)/2}
\pgfmathsetmacro{\wLay}{\xTr-\xAl+0.5}

\node[lay, minimum width=\wLay cm] (L) at (\xMid,\ylay) {decoder layer $\ell$};

% tokens, dots and frame on both rows
\foreach \yy in {\yin,\yout}{
  \foreach \i in {0,...,5,7,8}{ \node[aud] at (\i*\dx,\yy) {}; }
  \node[dts] at (6*\dx,\yy) {$\cdots$};
  \foreach \i in {9,11}{ \node[txt] at (\i*\dx+\gap,\yy) {}; }
  \node[dts] at (10*\dx+\gap,\yy) {$\cdots$};
  \draw[frm] (\xAl-0.2,\yy-0.24) rectangle (\xTr+0.2,\yy+0.24);
}
% input token -> layer, layer -> output token
\foreach \x in {0,1,2,3,4,5,7,8}{
  \draw[flw] (\x*\dx,\yin+0.18) -- (\x*\dx,\ylay-0.22);
  \draw[flw] (\x*\dx,\ylay+0.22) -- (\x*\dx,\yout-0.18);
}
\foreach \x in {9,11}{
  \draw[flw] (\x*\dx+\gap,\yin+0.18) -- (\x*\dx+\gap,\ylay-0.22);
  \draw[flw] (\x*\dx+\gap,\ylay+0.22) -- (\x*\dx+\gap,\yout-0.18);
}

% BI compares each position's input and output
\draw[cmp] (\xTr+0.27,\yin) to[bend right=50]
  node[midway, right, lbl] {$\mathrm{BI}^{\mathcal{A}\cup\mathcal{T}}_\ell$} (\xTr+0.27,\yout);

% row labels
\node[lbl, anchor=east] at (\xAl-0.3,\yout) {$\mathbf{h}^{(n)}_{\ell+1}$};
\node[lbl, anchor=east] at (\xAl-0.3,\yin)  {$\mathbf{h}^{(n)}_{\ell}$};

% group labels above output row
\node[lbl, orange!70!black] at ({(\xAl+\xAr)/2},\yout+0.42) {audio tokens};
\node[lbl, blue!60!black]   at ({(\xTl+\xTr)/2},\yout+0.42) {text tokens};

% braces
\draw[brc, orange!70!black] (\xAr+0.2,-0.36) -- (\xAl-0.2,-0.36)
  node[midway, below=4pt, lbl] {$\mathrm{BI}^{\mathcal{A}}_\ell$};
\draw[brc, blue!60!black]  (\xTr+0.2,-0.36) -- (\xTl-0.2,-0.36)
  node[midway, below=4pt, lbl] {$\SBIdec_\ell$};
\end{tikzpicture}}